# Phase diagram of iron-arsenide superconductors Ca(Fe$_{1-x}$Co$_x$)$_2$As$_2$ (0 ≤ x ≤ 0.2)

L. Harnagea[*], S. Singh[#], G. Friemel, N. Leps, D. Bombor, M. Abdel-Hafiez, A. U. B Wolter, C. Hess, R. Klingeler[§], G. Behr, S. Wurmehl, and B. Büchner

Leibniz-Institute for Solid State and Materials Research, (IFW)-Dresden, D-01171 Dresden, Germany
[#]Indian Institute of Science Education and Research (IISER), Pune, Maharashtra 411008, India
[§]Kirchhoff Institute for Physics, University of Heidelberg, D-69120 Heidelberg, Germany

Platelet-like single crystals of the Ca(Fe$_{1-x}$Co$_x$)$_2$As$_2$ series having lateral dimensions up to 15 mm and thickness up to 0.5 mm were obtained from the high temperature solution growth technique using Sn flux. Upon Co doping, the *c*-axis of the tetragonal unit cell decreases, while the *a*-axis shows a less significant variation. Pristine CaFe$_2$As$_2$ shows a combined spin-density-wave and structural transition near T = 166 K which gradually shifts to lower temperatures and splits with increasing Co-doping. Both transitions terminate abruptly at a critical Co-concentration of $x_c$ = 0.075. For x ≥ 0.05, superconductivity appears at low temperatures with a maximum transition temperature $T_C$ of around 20 K. The superconducting volume fraction increases with Co concentration up to x = 0.09 followed by a gradual decrease with further increase of the doping level. The electronic phase diagram of Ca(Fe$_{1-x}$Co$_x$)$_2$As$_2$ (0 ≤ x ≤ 0.2) series is constructed from the magnetization and electric resistivity data. We show that the low-temperature superconducting properties of Co-doped CaFe$_2$As$_2$ differ considerably from those of BaFe$_2$As$_2$ reported previously. These differences seem to be related to the extreme pressure sensitivity of CaFe$_2$As$_2$ relative to its Ba counterpart.

## I. INTRODUCTION

Iron-pnictides have attracted enormous attention since the discovery of superconductivity in fluorine doped LaOFeAs with a superconducting transition temperature ($T_C$) of 26 K.[1] Shortly after this discovery, higher $T_C$ values were achieved in these oxypnictides either by replacing La with smaller size rare-earth ions,[2-4] or by applying pressure,[5] elevating $T_C$ in this class of materials to as high as 55 K.[4] These compounds adopt a layered structure consisting of an alternating stack of LaO and FeAs layers perpendicular to the crystallographic *c*-axis. Since the original discovery, several other families of Fe-based layered high $T_C$ superconductors have also been discovered.[6-9] One of these families constitutes the ThCr$_2$Si$_2$ structure-type *A*Fe$_2$As$_2$ ('122' family of compounds, *A* = Ca, Sr, Ba, Eu) intermetallics. They adopt a layered structure analogous to oxypnictides with LaO layers replaced by layers of *A*. The pristine compounds are characterized as poor metals showing structural transitions in the temperature range 140 to 205 K,[10–14] and an antiferromagnetic ground state due to Spin Density Wave (SDW) formation of the Fe 3*d*-spins. Analogous to the oxypnictides, superconductivity can be induced in these compounds, with $T_C$ as high as 38 K, either by chemical substitution (electron/hole doping) or by external pressure.[6, 15-22] The electronic phase diagrams for Co-doped Ba-122 and Sr-122 compounds were reported based on single-crystalline studies.[23-26] In the Co-doped Ba-122 series, the electronic phase diagrams reported by several groups showed good agreement with each other,[23-25] which is an indication that reproducible single-crystals of these compounds can be grown for comprehensive investigations. More recently, superconductivity has also been reported in 4*d* and 5*d* transition metal doped Ba-122,[19] and Sr-122 compounds,[20] and their phase diagrams have also been mapped out. While the electron/hole doped $BaFe_2As_2$ and $SrFe_2As_2$ series of compounds have remained very much in focus, there are much scarce studies on the analogous $CaFe_2As_2$ compounds.

Recently, pristine $CaFe_2As_2$ has attracted a great deal of attention due to the extreme pressure sensitivity of its structural and magnetic behavior to external pressure. Upon application of a modest hydrostatic pressure of 0.35 GPa, the first order transition from the tetragonal (*T*) / paramagnetic state to the orthorhombic (*O*) / antiferromagnetic one is suppressed and above 0.35 GPa it completely vanishes being replaced by a first order transition to a low temperature collapsed tetragonal (*cT*) non-magnetic phase. Further application of pressure leads to an increase of the transition temperature from *T* to *cT* phase crossing room temperature around 1.7 GPa.[21, 27, 28]

It is noteworthy that this behavior depends crucially on the hydrostaticity of the pressure medium. In He - gas cells where hydrostatic conditions can be as close as possible realized, first – order like phase transitions were observed and at low temperatures the sample exhibits single crystallographic and magnetic phases.[29]

In contrast, the transition broadens and the sample enters into a low-temperature multi-crystallographic phase state when liquid media cells are used.[28] Interestingly, superconductivity at low temperature is observed only for the later case, indicating that the non-hydrostaticity of the pressure medium is an important ingredient in stabilizing a superconducting state in $CaFe_2As_2$ under pressure.[21]

The superconducting phase is characterized by a dome ($T_C \sim 12$ K) centered at 0.5 GPa, extending from 0.23 GPa up to 0.86 GPa.[21] These findings were further reinforced in Ref. [30] where superconductivity below 10 K was reported under uniaxial pressure.

Recently, it was shown that electronic doping through substituting of Fe for Co in $CaFe_2As_2$ also leads to the appearance of superconductivity at temperatures below about 20 K. More specifically, about 3% Co doping in Ca-122,[31] (4% in Ref. [32]) results in complete suppression of the magnetic/structural transition and emergence of superconductivity below 17 K (20.4 K in Ref. [32]).

However, a systematic study aimed at understanding the influence of progressive increase in cobalt doping level in Ca-122 is still lacking. The extreme sensitivity of $CaFe_2As_2$ on applied pressure suggests that in addition to changing the electronic doping level, the effect of chemical pressure upon Co substitution might be relevant too. Hence, one should expect a more complex influence of Co-doping on Ca-122 than in the other members of the 122 family, which motived us to take up the present studies.

Here, we present a complete doping dependence of magnetic, structural and superconducting properties of single-crystalline $Ca(Fe_{1-x}Co_x)_2As_2$ samples over a wide doping range from $x = 0$ to $x = 0.20$.

Experimental procedures and single crystal growth experiments are presented in section II. Single crystals were characterized using different techniques which are presented and discussed in section III. Finally, the resulting phase diagram of $Ca(Fe_{1-x}Co_x)_2As_2$ are shown and compared with the well studied case of $Ba(Fe_{1-x}Co_x)_2As_2$ which allowed us to make some general preliminary remarks regarding the role of electron doping vis-à-vis to chemical pressure due to Co doping in the 122 family.

## II. EXPERIMENTAL PROCEDURE

Single crystals of $Ca(Fe_{1-x}Co_x)_2As_2$ ($0 \leq x \leq 0.20$) were grown by employing the high temperature solution growth method with Sn flux. All room-temperature processing (weighing, mixing, grinding and storage) was carried out in an Ar filled glove-box ($O_2$ and moisture level less than 0.1 ppm). The growth experiments were carried out in three steps. In the first step, the precursor materials (CaAs, $Fe_2As$ and $Co_2As$) were prepared by reacting stoichiometric quantities of the constituents under vacuum at temperatures less than 900 ºC. $Fe_2As$ and $Co_2As$ are prepared by mixing, respectively, Fe (Alfa Aesar, 99.998%) and Co (Heraeus, 99.8%) powders with stoichiometric amount of arsenic lumps (Alfa Aesar, 99.999%) after removing the surface oxidation products by sublimation method. For synthesis of CaAs, Ca shots (Merck, 98.5%, 2-6 mm shots) and As lumps were physically separated in a quartz ampoule sealed under vacuum, slowly heated up to 850 ºC. This temperature was maintained for a period of 2 days allowing As vapours to slowly react with Ca to form CaAs. In the second step, stoichiometric quantities of the pre-reacted materials CaAs, $Co_2As$ and $Fe_2As$ were weighed, grounded and pressed into pellets and sealed in a quartz ampoule under vacuum. The ampoule was slowly heated to 750 ºC, kept there for 10 h and then further heated to 850 ºC for another 15 h before cooling to room temperature. In the final step, the sintered $Ca(Fe_{1-x}Co_x)_2As_2$ pellets and the Sn shots, taken in a molar ratio of 1 : 30, were placed in an alumina crucible together with a second catch crucible containing quartz wool and subsequently sealed in an quartz ampoule under vacuum. The charge was slowly heated to 1090 ºC, kept there for 24 h to ensure the homogenization and then cooled down to 600 ºC under a slow cooling rate of 2.5 ºC per hour. At this temperature, the Sn flux was decanted by flipping the ampoule up-side-down. After 3-4 h of waiting, the ampoule was cooled down to room temperature by switching off the furnace. We found that the three steps procedure described above prevents the growth of a competing orthorhombic phase of stoichiometry $CaFe_4As_3$ that can easily be identified due to its thin needle-like morphology contrary to the platelet-like single crystals of $CaFe_2As_2$ phase. The formation of this parasitic phase has also been recognized and reported earlier by other groups.[10, 33] However, using our protocol only platelet-like crystals of $CaFe_2As_2$ can be obtained as shown in Fig. 1(a). Figure 1 (b) shows the flux after the decantation. In Fig. 1 (c-e), a few representative pictures of as-grown single crystals on an mm-size grid are shown. All crystals are platelet-like, shiny, malleable and prone to exfoliation. Their lateral dimensions are as large as 15 mm and thickness ranging from 200 to 500 μm.

The lattice parameters of the all the grown crystals were determined by x-ray powder diffraction (XRD using a Rigaku Miniflex diffractometer (Cu Kα radiation)). For this purpose, few small single crystalline pieces from each batch were ground into a fine powder along with high-purity silicon powder added as an internal standard. The growth behaviour, crystal morphology and Co-composition were accessed using a Scanning Electron Microscope (SEM Philips XL 30) equipped for energy-dispersive x-ray spectroscopy probe (EDX) and wavelength dispersive x-ray probe (WDX). The temperature dependence of the electric resistivity from 4 to 296 K was measured using a standard four probe technique after cutting a platelet-like crystal in a rectangular parallelepiped shape whose largest surface is parallel to the crystallographic *ab*-plane. Current and voltage probes were made using copper wires glued to the sample surface (parallel to the *ab*-plane) using silver epoxy. The magnetization measurements were performed using a SQUID magnetometer (MPMS-XL5) from Quantum Design. Data have been obtained in the temperature range 2 to 50 K in a magnetic field of 2 mT and between 2 and 350 K in an applied magnetic field of 1 T.

## III. RESULTS AND DISCUSSIONS

### A. Structure analysis, composition, and the growth behaviour of Ca(Fe$_{1-x}$Co$_x$)$_2$As$_2$ single crystals

Several platelet-like single crystals from each growth experiment were examined in detail under a scanning electron microscope equipped with EDX and WDX probe. On each sample, the composition is determined using the EDX data averaged over sixteen to twenty five different spots. The EDX data plotted in the Fig. 2 show that the actual Co content of these crystals is generally smaller than the nominal value. We observed that the standard deviation in Co concentration measured over several samples of the same batch and about 16 to 25 different spots on a each sample is less than 0.3 atomic % which is within the error limit of the EDX technique (1-2 at. %). These results demonstrate a fairly homogeneous distribution of Co within a single crystal and several single crystals of the same batch. Since the precision of EDX is limited, henceforth, we will refer to our single crystals by their nominal Co composition ($x_N$), quoting their EDX value ($x_{EDX}$) wherever needed. A linear fit (dashed line in Fig. 2) of EDX compositions as a function of nominal Co concentrations exhibits a slope of about 0.7 which allows converting nominal values into EDX values and vice - versa.

It should be mentioned here that in the case of BaFe$_2$As$_2$ single-crystals grown using Sn-flux the presence of up to 1 at % of Sn incorporation in the structure has been reported.[15, 34] In order to exclude the incorporation of Sn in our crystals, we carried out WDX analysis at several points of the single crystals and repeated this experiment for several pieces from different batches. However, no trace of Sn in our single crystals could be detected using WDX analysis. Moreover, the resistivity and magnetization data of our undoped single crystal agree with those reported in the literature for both Sn and self-flux grown single crystals,[16, 33] unlike the case of Ba122, where the behaviour of Sn grown crystals markedly differ from those of self-flux grown and polycrystalline samples.[11, 17, 34] Therefore, we can exclude any Sn incorporation in our single crystals. The difference in growth behaviour between Ba122 and Ca122 could be related to the smaller ionic size of Ca compared to that of Ba.

Figure 3 shows representative powder diffraction patterns of the end members Ca(Fe$_{1-x}$Co$_x$)$_2$As$_2$ ($x_N$ = 0, 0.20). The observed powder pattern is indexed based on ThCr$_2$Si$_2$-type tetragonal structure (space group: I4/mmm, No. 139). Generally, the samples proved to be single phase, only arbitrarily we observed the diffraction lines of Sn due to the residual flux sticking on the surface of the single crystals. The lattice parameters of the parent compound obtained from the powder pattern are $a_0$ = 3.883 Å and $c_0$ = 11.750 Å. These values are in good agreement with those previously reported in the literature.[16, 35] Figure 4 shows the lattice parameter variation of the Co doping series as a function of increasing Co content. While the *c*-parameter of the tetragonal unit cell decreases upon Co-doping, the area of the *ab*-plane shows only a slight decrease. The general trend in the variation of the cell parameters in Ca(Fe$_{1-x}$Co$_x$)$_2$As$_2$ is quite similar to that observed in the Co-doped Ba-122 series of compounds.[25] However, the changes of the *c*-axis upon Co doping are more pronounced in the Ca-122 than in the Ba-122 series. Additionally, an x-ray diffraction experiment was also performed by mounting a single-crystalline platelet in Bragg-Brentano geometry. In this case the diffraction pattern showed only peaks with Miller indices [0, 0, 2*l*] (l = 1, 2, 3…) which implies that the *c*-axis is perpendicular to the plane of the platelet (Fig. 5).

We have investigated the Ca(Fe$_{1-x}$Co$_x$)$_2$As$_2$ single-crystals habits, morphology and the growth behaviour by electron microscopy. Few selected SEM images of Ca(Fe$_{1-x}$Co$_x$)$_2$As$_2$ single crystals are shown in Fig. 6. These crystals grow as layered structure, exhibit a platelet-like morphology and are easy to cleave along the *ab*-plane. The crystals are malleable and the layers present a high tendency to exfoliate as shown in Fig. 6 (a), (b) and (d). Figure 6 (b) shows that the crystals present macro-steps (or micro – steps) in the plane of the platelet, indicating a layer by layer growth mechanism due to step propagation. Occasionally, we observed the presence of step bunches on the plane of the platelets (Fig. 6 (c)). Here, the growth seems to be hindered due to surface Sn inclusions. In the same way, Sn precipitates can be trapped as inclusions (Fig. 6 (d)).

### B. Normal and Superconducting state properties

The temperature dependence of the in-plane resistivity ρ(T) of Ca(Fe$_{1-x}$Co$_x$)$_2$As$_2$ single crystals is shown in Fig. 7. The data for each sample are normalized to the value of resistivity at T = 296 K. The resistivity data for the pristine CaFe$_2$As$_2$ compound exhibits a metallic behaviour over the entire temperature range with a prominent first-order anomaly at T$_0$ ~ 166 K, where the resistivity shows a step-like increase upon cooling (Fig. 7 (a) - 8 (a)). The position and shape of the anomaly is in agreement with the data previously reported. We should note that the temperature where the anomaly is observed can range between 165 and 172 K,[10, 16, 35] depending on the growth method. Upon cooling below $T_0$, the crystal symmetry lowers from tetragonal to orthorhombic; concomitantly, the Fe moments order antiferromagnetically (AFM).[10, 36] The temperature variation of resistivity also shows a hysteresis of about 2 K between the cooling and the warming run at the combined structural/magnetic transition (not shown) consistent with previous transport studies.[10, 32]

Upon Co substitution, the sharp first-order structural/magnetic anomaly of the parent compound gradually broadens and shifts to lower temperatures (Fig. 7 (a)-(d) and Fig.8 (a)–(d)) and it is completely suppressed for Co concentrations larger than 0.075. Figure 8 shows the splitting of the structural and magnetic transitions for the intermediate values of Co concentration. We mention that our thermal expansion data on $Ca(Fe_{0.925}Co_{0.075})_2As_2$ confirm clearly the presence of two distinct transitions.[37]

We assigned the first transition to the strong change of the slope of $d\rho/dT$ while the second transition was defined as the inflection point of the $d\rho/dT$ curve in line with Ref. 25 as demonstrate in Fig. 8.

The first derivative of resistivity ($d\rho/dT$) for $x_N = 0.03$ shows no distinct difference between the structural and the magnetic transition (Fig. 8 (b)). However, the temperature interval of the transition is broader compared to the undoped sample and shifted by about 17 K to lower temperatures with respect to the parent compound. For $x_N = 0.05$ and 0.075 the combined structural/magnetic anomaly of the pristine compound actually splits into two distinct anomalies at 142 K, 131 K and 109 K, 89 K, for $x_N = 0.05$ and 0.075, respectively (Fig. 8 (c) and (d)). The error in the determination of the structural ($T_S$) and magnetic ($T_N$) transitions temperatures can be estimated at around 3 K if we take into account that the peak in the first derivate of the resistivity is relatively broad (Fig. 8 (b) – (d)). Based on a recent neutron study in analogous Ba-122 compounds,[38] we have tentatively attributed the lower temperature to the onset of antiferromagnetic order and the higher temperature to the occurrence of the structural phase transition. Neutron studies on Co-doped Ca-122 samples would be helpful in addressing this point further in case of the Ca series.

In the following, we will discuss the resistivity behaviour of the samples which presents a partial or complete drop to zero resistivity at low temperatures as shown in Fig. 9 where the normalized resistivity, $\rho(T)/\rho(30 K)$, is shown below T = 30 K. Interestingly, already the sample with the lowest Co concentration in our studies ($x_N = 0.03$) shows a slight decrease of the resistivity at low temperature which can be associated with the onset of spurious superconductivity below about 15 K.

With increasing Co-concentration the superconducting drop gets more pronounced and shifts to higher temperature, however, the resistivity remains finite down to T = 4 K at $x_N$ = 0.05. In contrast, zero resistivity at low temperatures is observed for Co concentrations above and including 0.075 and up to 0.135. $T_C$ from the resistivity curve, $T_C^\rho$ (determined using the 90% criterion; i.e., the temperature where resistivity has decreased by 90 % of its normal state value), is estimated to be 18.3 K, 15.5 K, 17.6 K, 8 K and 7.9 K for $x_N$ = 0.075, 0.09, 0.10, 0.125, 0.135. For higher Co concentration $T_C$ decreases further and the superconducting drop turns partial as for the underdoped samples. Thus, at first sight the evolution of superconductivity and the suppression of the high-temperature structural/magnetic transition in $Ca(Fe_{1-x}Co_x)_2As_2$ seems to follow a trend analogous to the $Ba(Fe_{1-x}Co_x)_2As_2$ compounds. However, in the present case, $T_C$ attains its maximum value ($x_N$ = 0.075) when the structural/magnetic anomaly is still present at relatively high temperatures close to T = 100 K and upon further increase of the Co-concentration the structural/magnetic anomaly vanishes abruptly, while $T_C$ changes only marginally. The superconducting transition width ($\Delta T_C^\rho$) for these single-crystals is estimated using: $\Delta T_C^\rho$ = [T(90 %) – T(10 %)], where T (90 %) and T (10 %) are, respectively, the temperatures where ρ(T) is 90 % and 10 % of its normal state value. The value of $\Delta T_C^\rho$ is around 3 K for $x_N$ between 0.075 and 0.10 and higher than 5 K for $x_N$ = 0.125, 0.135. These transition widths are very large compared to the analogous Ba compounds. For example, $\Delta T_C^\rho$ for an optimally Co-doped Ba122 sample is about 0.7 K.[17, 39]

We shall enquire into these differences in details later after having presented the other salient features of our resistivity and magnetization data. However, it should be remarked here that these samples exhibit a fairly homogeneous Co distribution within the limits of the EDX analysis as discussed in the previous section.

We briefly remark on the residual resistivity ratio (RRR) of $Ca(Fe_{1-x}Co_x)_2As_2$ series taken as RRR = ρ(296K)/ρ(25K). The parent compound has a resistivity of ~ 3.7 μΩ·m at 296 K and it presents a RRR of about 4 which is in a good agreement with the data previously reported.[16, 31] For the samples with less then 5% Co content RRR is ~ 2 and increases with further increase in the doping level up to 21.3 at 10% Co content which is then followed by a decrease to 11.5 at $x_N$ = 0.2. The high RRR values are usually interpreted as a signature of a high quality of the single crystal. However, we should note that in the case of these SDW ordering compounds, where parts of the Fermi surface are gapped in the SDW state, it is uncertain to conclude a higher crystalline quality from a large RRR value. The actual value of the resistivity at low temperature in such cases is determine by a delicate balance between a reduced carrier density (caused by the gap) and an enhanced carrier mean free path (due to reduced scattering at low temperatures) both of which depend differently on the sample purity.

We turn now to the magnetic properties of our samples. Figure 10 shows the temperature dependence of M/H of Ca(Fe$_{1-x}$Co$_x$)$_2$As$_2$ single crystals measured under zero-field-cooled (ZFC) conditions in a magnetic field of 1 T applied along the *ab*-plane.[1] The pristine CaFe$_2$As$_2$ compound exhibits a sharp anomaly at 166 K due to the combined structural/magnetic transition consistent with the resistivity data discussed above. The overall behaviour of magnetization in the pristine compound is similar to that reported previously.[35] The increase in magnetization at temperatures below 25 K is reminiscent of a Curie-like tail due to magnetic impurities, which is often observed in compounds with small net magnetization in their antiferromagnetic ground state as in the present case. The magnetization data of the Co-doped samples correlate nicely with their resistivity data shown in Fig. 7. With increasing Co-concentration, the anomaly associated with the structural/magnetic phase transition gradually shifts to lower temperatures and disappears completely, and rather abruptly, for $x_N \geq 0.075$. The first derivatives of the magnetization curves corresponding to $x_N = 0.05$ and $x_N = 0.075$ show splitting of the structural/magnetic anomaly into two distinct anomalies at $T_S \sim 141$ K, $T_N \sim 131$ K and $T_S \sim 108$ K, $T_N \sim 92$ K, respectively, which is in fair agreement with the analysis of the resistivity data.

The drop in the M/H data of Co-doped samples at low temperatures (Fig. 10) is due to the onset of superconductivity, which is also studied by measurements of the zero-field-cooled volume susceptibility ($\chi_v$) under an applied field of 20 Oe parallel to the *ab*-plane of the crystal (Fig. 11). Since the full diamagnetic screening corresponds to $4\pi\chi_v = -1$, the magnitude of the diamagnetic susceptibility, $4\pi\chi_v$, represents the superconducting volume fraction of the sample. As can be seen in the figure, the superconducting volume fraction significantly varies with doping concentration. This variation (Fig. 11) demonstrates that the superconductivity is complete only within a very narrow range of the Co concentration. Interestingly, the sample for which $T_C$ is maximum ($x_N = 0.075$) exhibits only a partial superconducting volume fraction. The superconducting transition temperature from the susceptibility plot ($T_C^\chi$) is taken at the intersection point of the slopes extrapolated from the normal state and the superconducting transition region, respectively. Using this procedure yields $T_C$ of 18.5 K, 15 K, 17 K, 14.1 K, 11.8 K and 11.4 K for samples with $x_N$ = 0.075, 0.09, 0.10, 0.125, 0.135, 0.15.

---

[1] A few of the Ca(Fe$_{1-x}$Co$_x$)$_2$As$_2$ single crystals used in the present work have also been investigated in References 40 and 41.

Using this criterion for determining $T_C$, we observed a good agreement between the transition temperature values determined from resistivity and magnetization data for Co doping up to 0.10. On the other hand, a large difference up to 6 K in the values of the transition temperature determined from resistivity and magnetization curves for the samples from the over-doped regime was observed.

However, if we use the onset criterion for extracting the transition temperature from the resistivity data a fairly agreement is observed between $T_C$ values. Note as well that, in contrast, applying an onset criterion for extracting the transition temperature from magnetization data yields only a weakly doping dependent $T_C$ value of around 20 K for all doping levels.

## C. Phase diagram

Using the magnetic and the electrical resistivity data we traced the evolution of the structural / magnetic phase transition and the superconducting phase transition of $Ca(Fe_{1-x}Co_x)_2As_2$ single crystals as a function of Co content (Fig. 12). In the under-doped side of the phase diagram ($0.05 \leq x_N \leq 0.075$), there is a small region which shows the apparent coexistence of the orthorhombic/antiferromagnetic phase with superconductivity. This coexistence is particularly remarkable for the sample with $x_N$ = 0.075 for which the $T_C$ is maximum and yet the structural/magnetic anomalies occur at relatively high temperatures close to 100 K. Upon further increasing of the Co-concentration the anomaly is no longer visible in the magnetic and resistivity data, while superconductivity remains, albeit at slightly lower temperatures. Concomitantly, the superconducting volume fraction increases to full diamagnetic screening at $x_N$ = 0.09 (Fig. 13). At $x_N \geq 0.10$ both $T_C$ and the superconducting volume fraction decrease again.

Even though the superconducting region has a usual dome-like appearance, the dome maximum appears at the under-doped side of the phase diagram in contrast to the $Ba(Fe_{1-x}Co_x)_2As_2$ phase diagram,[25] where the dome-maximum appears at optimal doping. In addition, suppression of the structural/antiferromagnetic phase transition upon Co doping is quite different in Ba and Ca compounds, too. In Ba(Fe$_{1-x}$Co$_x$)$_2$As$_2$ the structural/antiferromagnetic transition is monotonically and continuously suppressed with increasing Co substitution at an initial rate of roughly 15 K per atomic percent Co and disappears above approximately 5.8 % Co substitution. In the case of Ca(Fe$_{1-x}$Co$_x$)$_2$As$_2$, on the other hand, the structural/magnetic anomaly is suppressed initially at a rate of approximate 10 K per atomic percent Co but eventually vanishes abruptly above $x_N = 0.075$. Nonetheless, similar to the case of Co doped Ba–122, in Ca(Fe$_{1-x}$Co$_x$)$_2$As$_2$ the under-doped region exhibits a splitting of the structural and magnetic phase transitions.

In Fig. 13 we present in detail (a) the variation of the superconducting transitions temperatures as determined from susceptibility and resistivity data, (b) the superconducting volume fraction $V_{SC}$, (c) the transition width $\Delta T_C$, and (d) the residual resistivity ratio as a function of the nominal Co concentration of the samples. The data imply that both the superconducting volume fraction and the transition width exhibit drastically changes as a function of Co concentration. Interestingly, there is an *inverse* correlation between the transition width and the superconducting volume fraction. In the range of 7.5 - 10% Co content, $V_{SC}$ is increasing while the superconducting transition is becoming sharper. The data show that the sharpest superconducting transitions are associated with bulk superconductivity while the broader transitions are associated with only partial volume superconductivity.[2]

## IV. DISCUSSION

In order to understand the salient differences between the physical properties of Ca-122 and Ba-122 it appears reasonable to invoke chemical pressure effects in addition to pure electronic doping effects. This can be inferred from the decrease in lattice parameters upon Co-doping as shown in Fig. 4. Since Co doping has only little effect on the crystallographic *a*-axis, the chemical pressure is essentially uniaxial. Such a chemical pressure also appears in the analogous Ba and Sr compounds; however, extreme sensitivity of CaFe$_2$As$_2$ in particular to applied uniaxial pressure complicates the present scenario. Therefore, in Ca-122 several competing factors come into play with increasing Co-doping. More specifically, Co-substitution for Fe not only results in electron doping (prevalent in Ba compounds), but also induce a chemical pressure resulting in competing states at low temperatures that include the orthorhombic (*O*), the collapsed tetragonal (*cT*) and the residual untransformed tetragonal phase (*T'*).[30] This conjecture is based on recent uniaxial pressure experiments which stabilized the *cT* and *T'* phases at the expense of the orthorhombic phase.[30] Furthermore it was shown in this study that pressure induced superconductivity below T = 10 K in pristine CaFe$_2$As$_2$ arises due to the *T'* phase stabilized under uniaxial pressure applied along the *c*-axis, while the *cT* phase does not support superconductivity.

---

[2] We note that before our measurements of the magnetization in small magnetic fields of 20 Oe we applied a demagnetization sequence in order to diminish the remanent field. From different demagnetization sequences and after applying the so-called the ultra low-field option we estimate an error in determining $V_{SC}$ due to the remaining remanent field up to around 10% under our working conditions.

Our data do not allow separating the effects of chemical pressure and electronic doping, but one can qualitatively understand the differences of the Ca-122 phase diagram to its Ba- and Sr-counterparts. In the under-doped region of the phase diagram (i.e. for $x_N$ = 0.03, 0.05, 0.075), the onset of superconductivity at a relatively high temperature of about 17 K (compared to the pressure induced T$_C$ of 10 K in pristine CaFe$_2$As$_2$) most likely arises due to the gradual filling-up of the hole pockets due to Co-doping which lowers the antiferromagnetic transition and apparently favors superconductivity analogous to the effect of Co-doping in BaFe$_2$As$_2$. Note, however, that the superconducting onset is significantly higher in temperature than in Co doped Ba-122.[25] In further contrast to Ba-122, upon increasing the Co-doping level the signatures of the orthorhombic phase and the SDW order disappear abruptly at $x_N \approx 0.075$ which is reminiscent of a similar behavior observed in the fluorine doped LaOFeAs superconductors.[42] For $x_N \leq 0.075$, the orthorhombic phase and superconductivity coexists with SDW ordering like in Co doped Ba-122 and Sr-122. Whether SC coexists with magnetism in the same spatial domains or if there is nanoscopic phase segregation between these phases is a question, however, that needs to be pursued further.

The complete suppression of the *T-O* transition at $x_N$ > 0.075 is accompanied by an enhanced superconducting volume fraction and a sharpening of $\Delta T_C$ (cf. Fig. 13) up to $x_N \sim$ 0.1. In analogy with the uniaxial pressures study,[30] one may speculate that the suppressed orthorhombic phase and the enhanced superconducting properties in this intermediate region of the phase diagram are stabilized by the presence of the *T'* phase.

In the overdoped region of the phase diagram ($x_N$ > 0.1), superconductivity is apparently weakened (Fig. 13). Remarkably, the suppression is significantly more rapid than in the Ba-122 case. A plausible scenario to explain this difference is an increasing fraction of the *cT* phase at low temperatures. In this scenario, upon increasing the Co concentration the weight fraction of the *T′* phase decreases in favor of the *cT* phase as shown in the uniaxial pressure study.[30] This conjecture is corroborated by a rough estimate of the chemical pressure corresponding to a given Co-concentration which can be obtained by comparing the lattice parameters under pressure at T = 300 K given in Ref. [28] to the ambient variation of lattice parameter as a function of Co-doping shown in Fig. 4. We find that $x_N$ = 0.15 (i.e., for a slightly overdoped compound) translates to an external uniaxial pressure of about 0.5 GPa applied parallel to the *c*-axis. At this uniaxial external pressure, the transition from *T* to *cT* phase would appear at around T = 120 K (cf. the T-p phase diagram in Ref. [28]). Since it is argued that the *cT* phase does not exhibit superconductivity, the pressure effect hence at least qualitatively implies the presence of a competing phase and, consequently, a more rapid suppression of superconductivity than expected by pure electronic doping effects.

## IV. SUMMARY AND CONCLUSIONS

We have successfully grown large and high-quality single crystals of $Ca(Fe_{1-x}Co_x)_2As_2$ (0 ≤ x ≤ 0.20) from Sn-flux using the high temperature solution growth technique. All the grown crystals were found to be phase-pure crystallizing in a tetragonal $ThCr_2Si_2$-type structure. Upon Co doping, the area of the *ab*–plane only slightly varies, while the crystallographic *c*-axis of the tetragonal unit cell decreases at a rate which is even more pronounced than in the case of Co doped Ba-122. A detailed WDX/EDX analysis of these single crystals indicates a fairly homogenous Co distribution with no traces of Sn incorporation in the structure. The actual Co concentration in our samples is found to be smaller than the nominal one ($x_{EDX}$ ≈ 0.7 $x_N$). The magnetic and transport data of the parent compound manifest a clear first order transition (SDW/structural transition) near 166 K, in good agreement with previous reports. Upon doping the combined SDW/structural transition at 166 K is gradually suppressed and splits into two distinct transitions, giving way to superconductivity with a maximum $T_C$ near 20 K. From the magnetic and resistivity data we established the electronic phase diagram of $Ca(Fe_{1-x}Co_x)_2As_2$ (0 ≤ x ≤ 0.20). Interestingly, the maximum $T_C$ value is observed for 7.5% Co content at which magnetism and superconductivity still coexist. Upon increasing the doping, a small interval with enhanced superconducting properties in the absence of the orthorhombic/SDW phase is observed up to ~10% doping level. Further increase of the doping leads to a rather rapid suppression of superconductivity. Our findings suggest that both steric effect due to the chemical pressure as well as electronic doping effects need to be taken into account for understanding the peculiar appearance and disappearance of superconductivity in Ca122 series of compounds, which differs from their Ba and Sr analogues.

ACKNOWLEDGEMENTS

This work was supported by the Deutsche Forschungsgemeinschaft through Grant No. BE1749/12, and the Priority Programme SPP1458 (Grants No. BE1749/13 and GR3330/2). We thank M. Deutchmann, S. Müller-Litvanyi, R. Müller, J. Werner, S. Pichl and S. Gaβ for tehnical support.

* l.harnagea@ifw-dresden.de

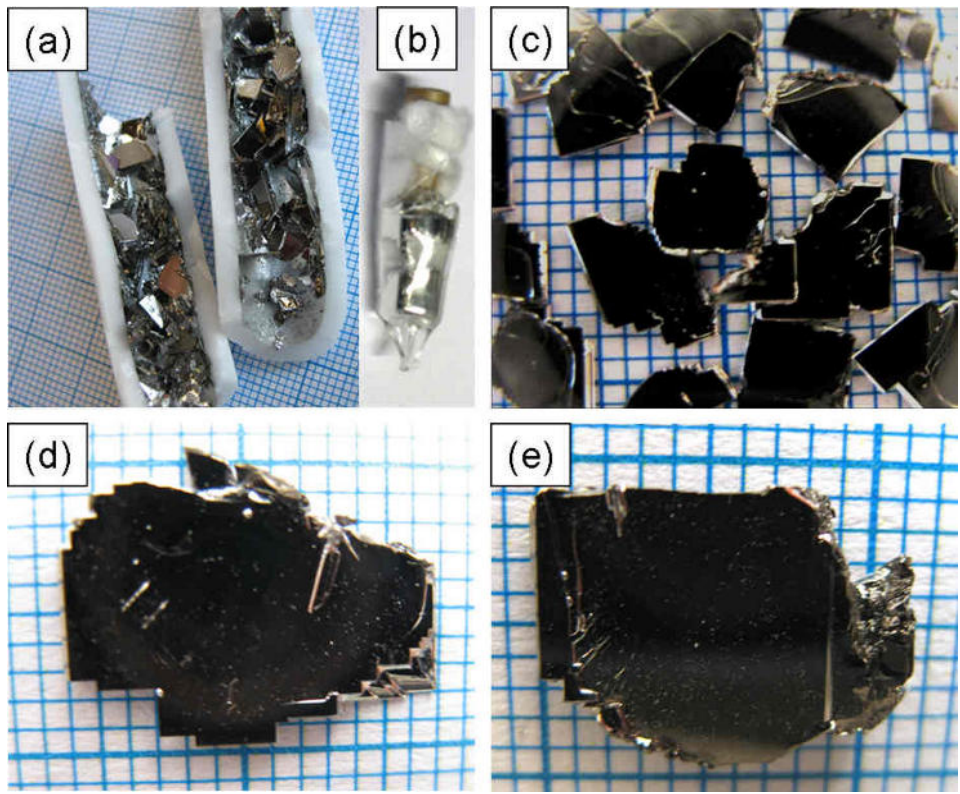

FIG. 1. (Color online) (a) As grown platelet-like single-crystals of Ca(Fe$_{1-x}$Co$_x$)$_2$As$_2$ after flux decanting. (b) Sn-flux separated from the crystals. (c) - (e) Typical flux grown single crystals of Ca(Fe$_{1-x}$Co$_x$)$_2$As$_2$ ($x_N$ = 0.05, 0.09, 0.15). The grid in each crystal is mm size.

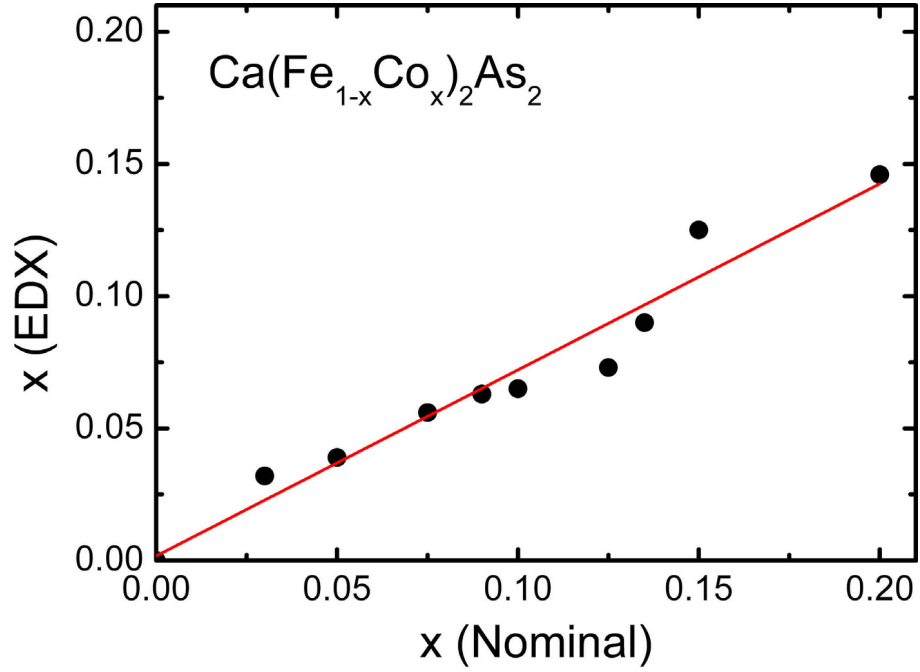

FIG. 2. (Color online) Co composition determined by means of EDX of Ca(Fe$_{1-x}$Co$_x$)$_2$As$_2$ single crystals as a function of nominal concentration x. The dashed line is a linear fit to the data with a slope of 0.7 (see text for details).

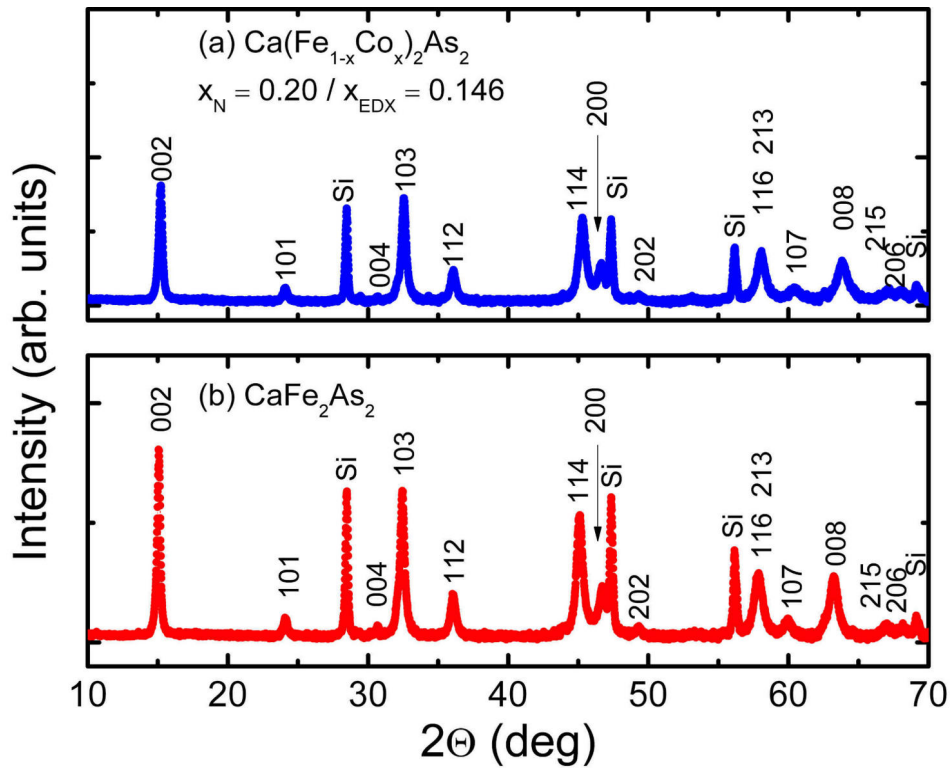

FIG. 3. (Color online) Powder x-ray diffraction pattern of Ca(Fe$_{1-x}$Co$_x$)$_2$As$_2$ (x$_N$ = 0, 0.20) single crystals. Peaks marked Si are due to silicon added as internal standard.

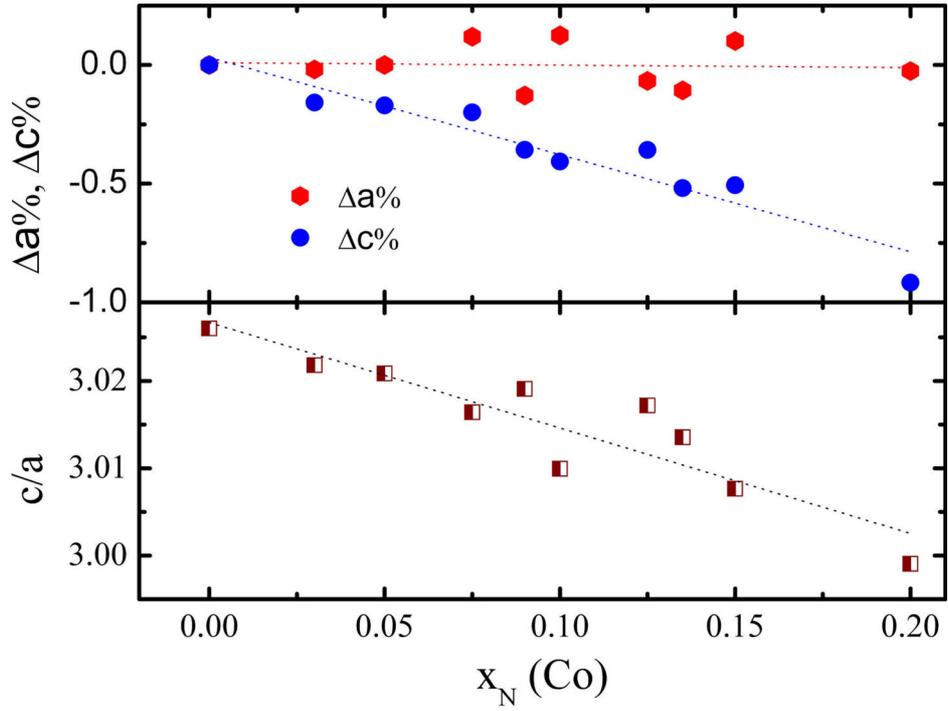

FIG. 4. (Color online) Variation of the lattice parameters as a function of nominal Co content. The lines are guides to the eye.

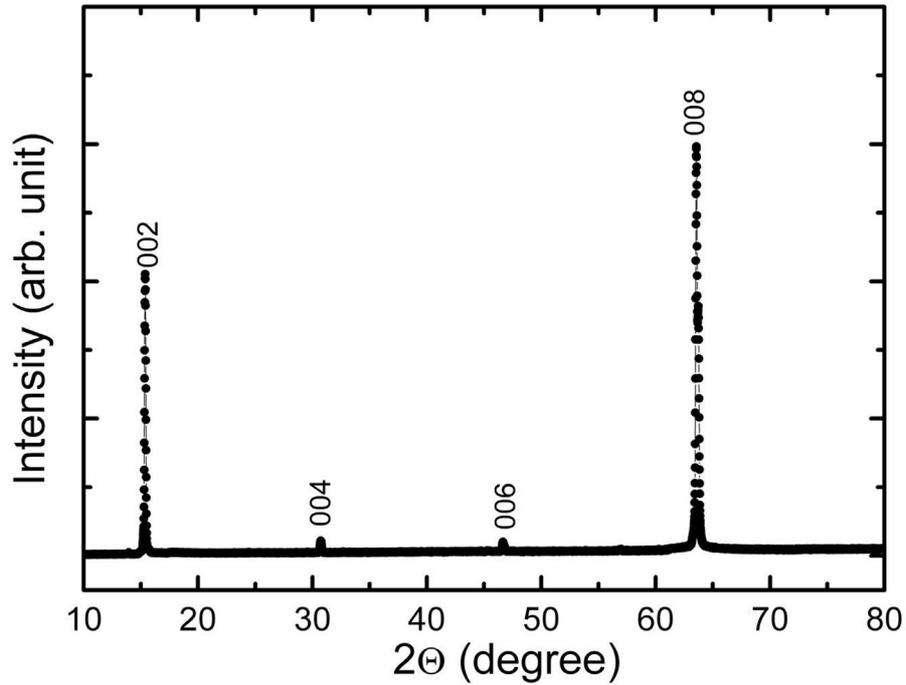

FIG. 5. (Color online) X-ray diffraction pattern due to a single crystalline CaFe$_2$As$_2$ in Bragg-Brentano geometry showing only [0 0 2l] reflections (see text for details).

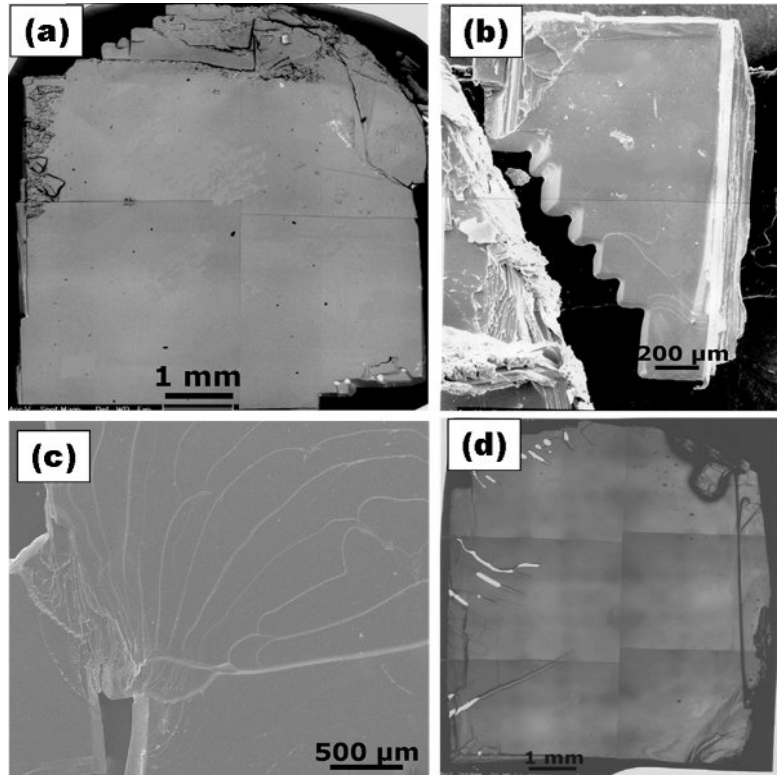

FIG. 6. (Color online) Representative SEM images of as-grown single crystals of Ca(Fe$_{1-x}$Co$_x$)$_2$As$_2$ (see text for details).

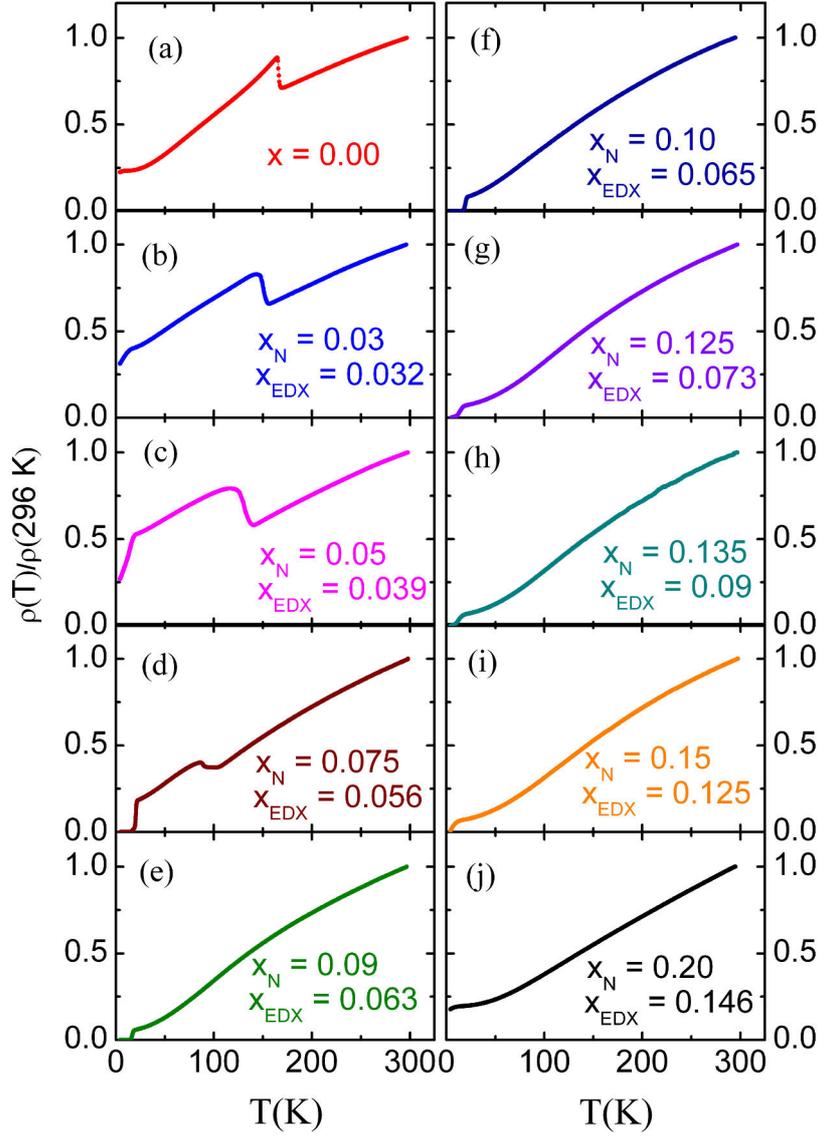

FIG. 7. (Color online) Temperature dependence of the in-plane electrical resistivity of Ca(Fe$_{1-x}$Co$_x$)$_2$As$_2$ single crystals, normalized to the resistivity value at 296 K. $x_N$ and $x_{EDX}$ are nominal and EDX composition, respectively.

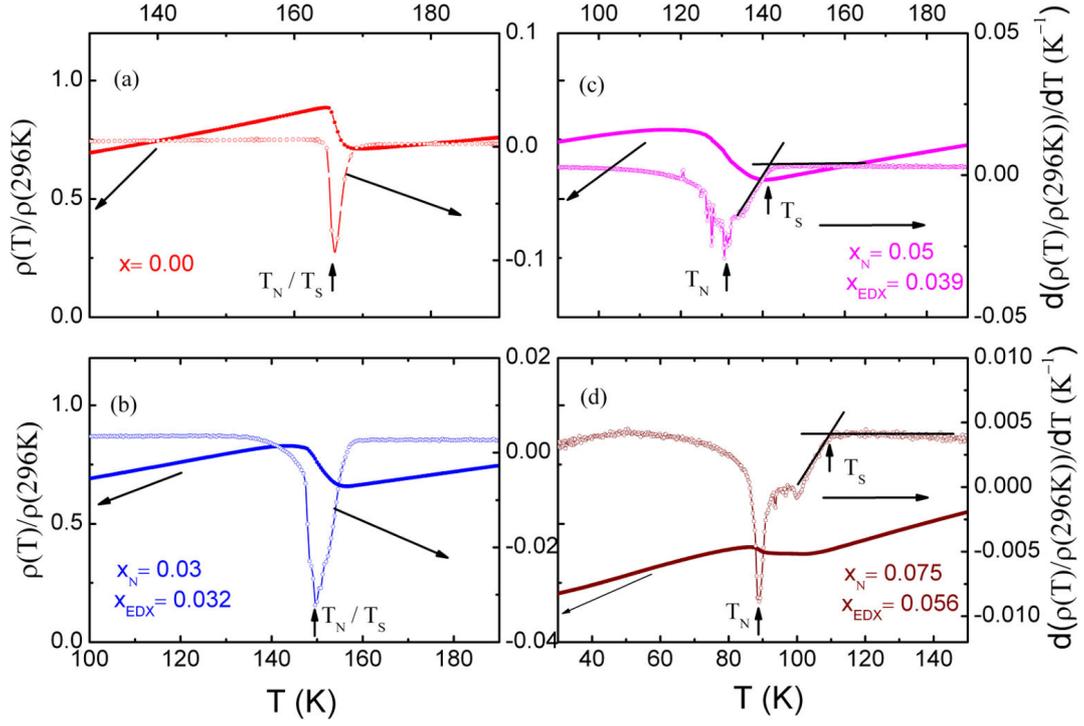

FIG. 8. (Color online) Normalized resistivity $\rho(T)/\rho(296K)$ of $Ca(Fe_{1-x}Co_x)_2As_2$ ($x_N$ = 0.00, 0.03, 0.05, 0.075) and its first derivate plotted as of the temperature. The arrows are indicating the magnetic ($T_N$) and the structural transition ($T_S$).

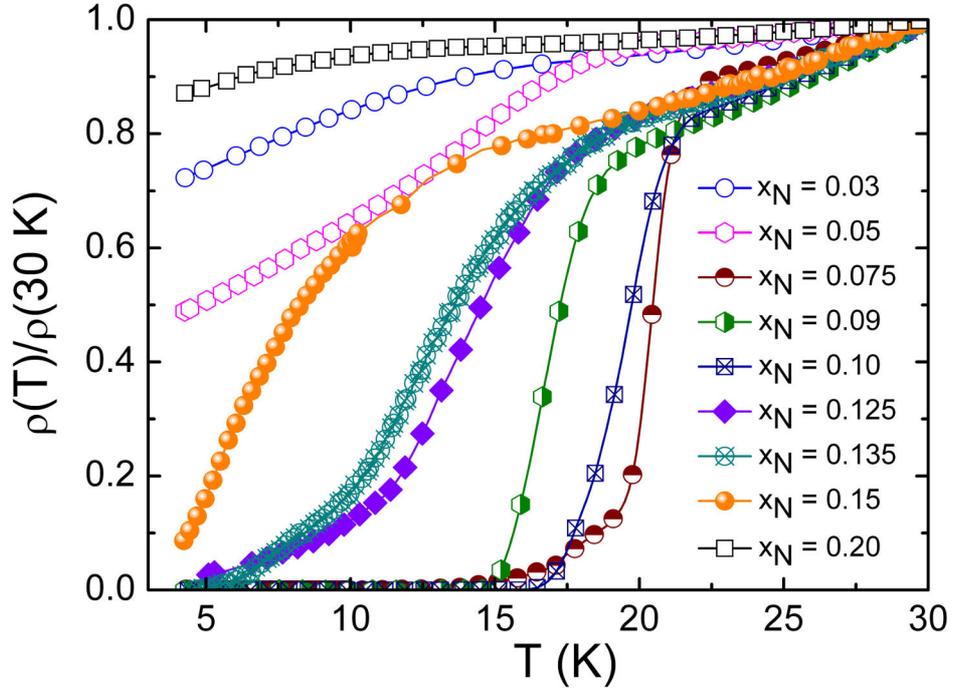

FIG. 9. (Color online) Low temperature resistivity $\rho(T)/\rho(30K)$ of $Ca(Fe_{1-x}Co_x)_2As_2$ showing the evolution of the superconducting transition with Co concentration.

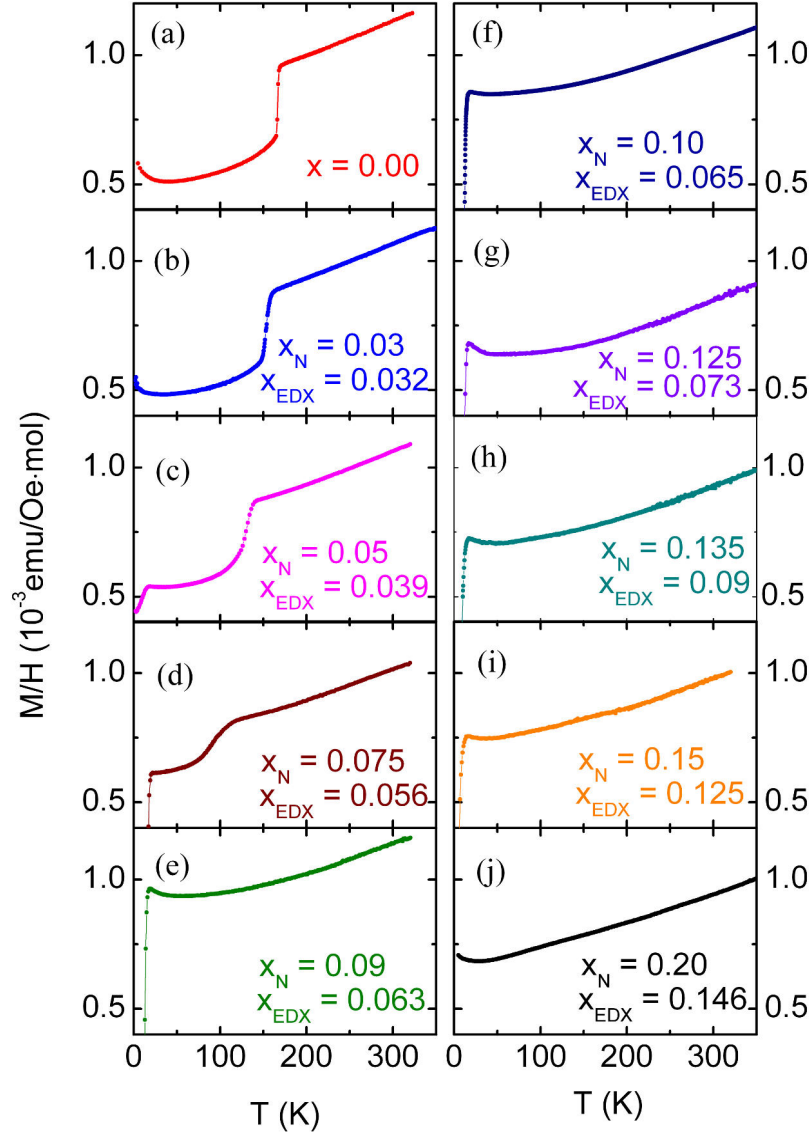

FIG. 10. (Color online) Temperature dependence of the magnetization of Ca(Fe$_{1-x}$Co$_x$)$_2$As$_2$ measured under an applied field of 1 T parallel to the crystallographic basal plane in zero-field conditions.

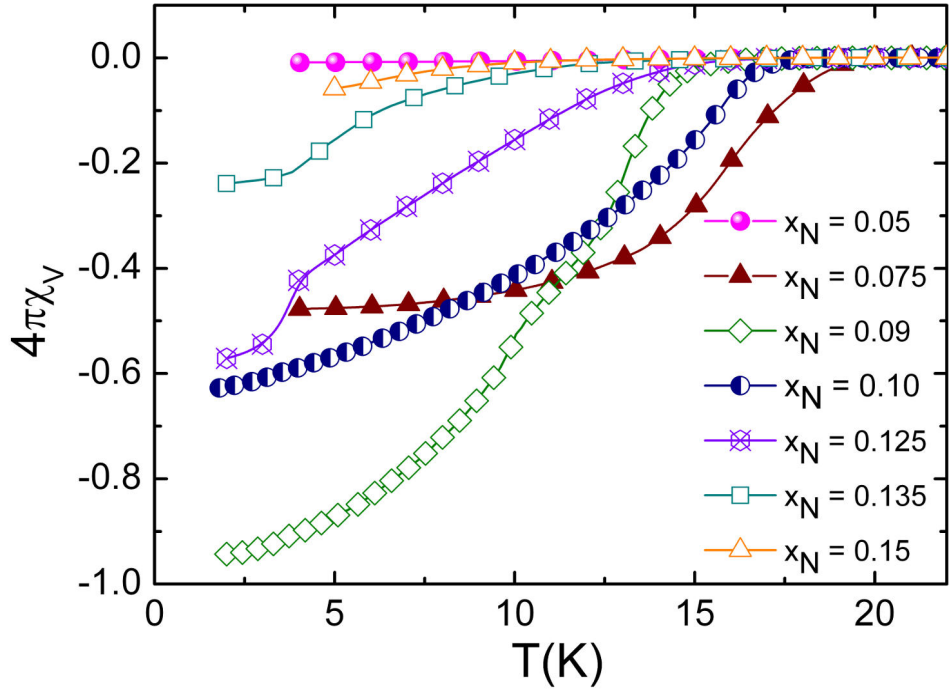

FIG.11. (Color online) Low temperature volume magnetic susceptibility of $Ca(Fe_{1-x}Co_x)_2As_2$ single crystals under 20 Oe zero-field cooled conditions parallel to the crystallographic basal plane.

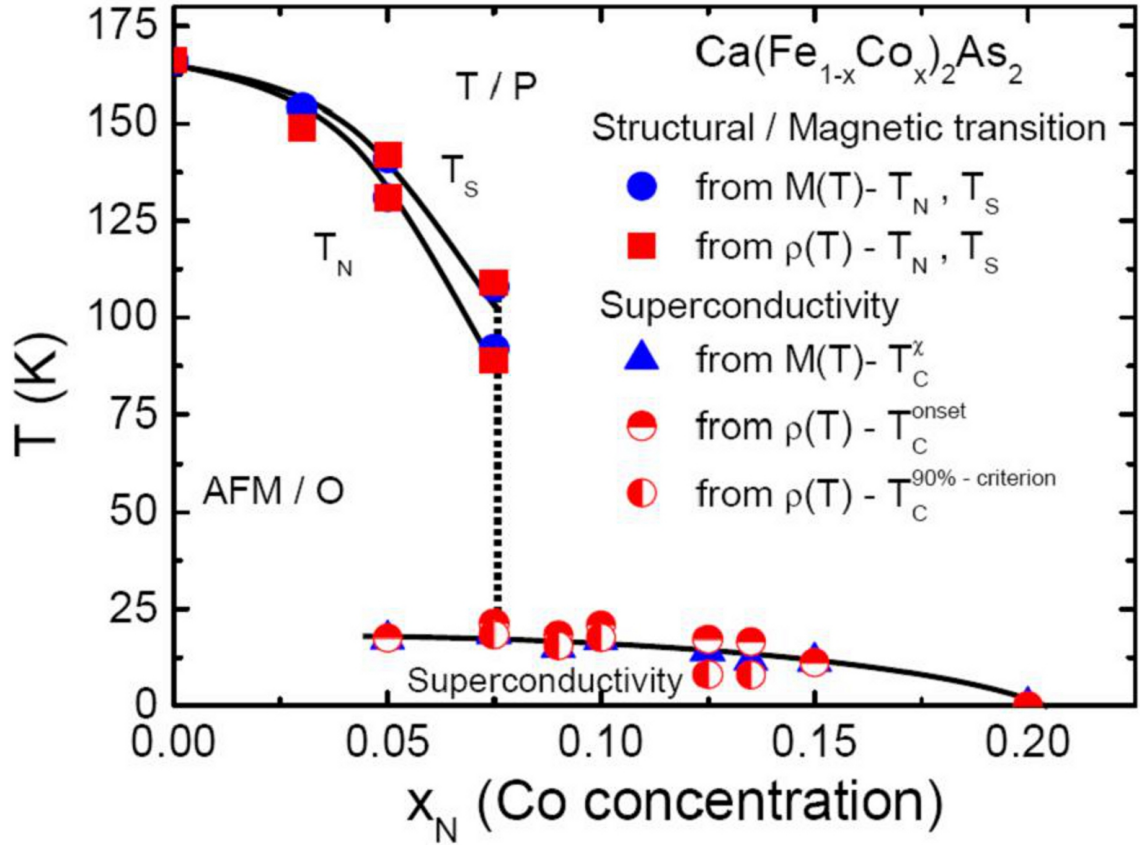

FIG. 12. (Color online) Electronic phase diagram of Ca(Fe$_{1-x}$Co$_x$)$_2$As$_2$ obtained from magnetic and electric resistivity data, showing the suppression of the magnetic/structural phase transition (blue circles and red squares – $T_N$, $T_S$) with increasing Co concentration and the appearance of the superconducting transition (red half filled circles - $T_C$ determined from resistivity using the 90% and the onset criterion; blue triangles - $T_C$ determined from magnetization data taken at the intersection point of the slopes extrapolated from the normal state and superconducting transition region) with maximum $T_C$ of ~ 20 K.

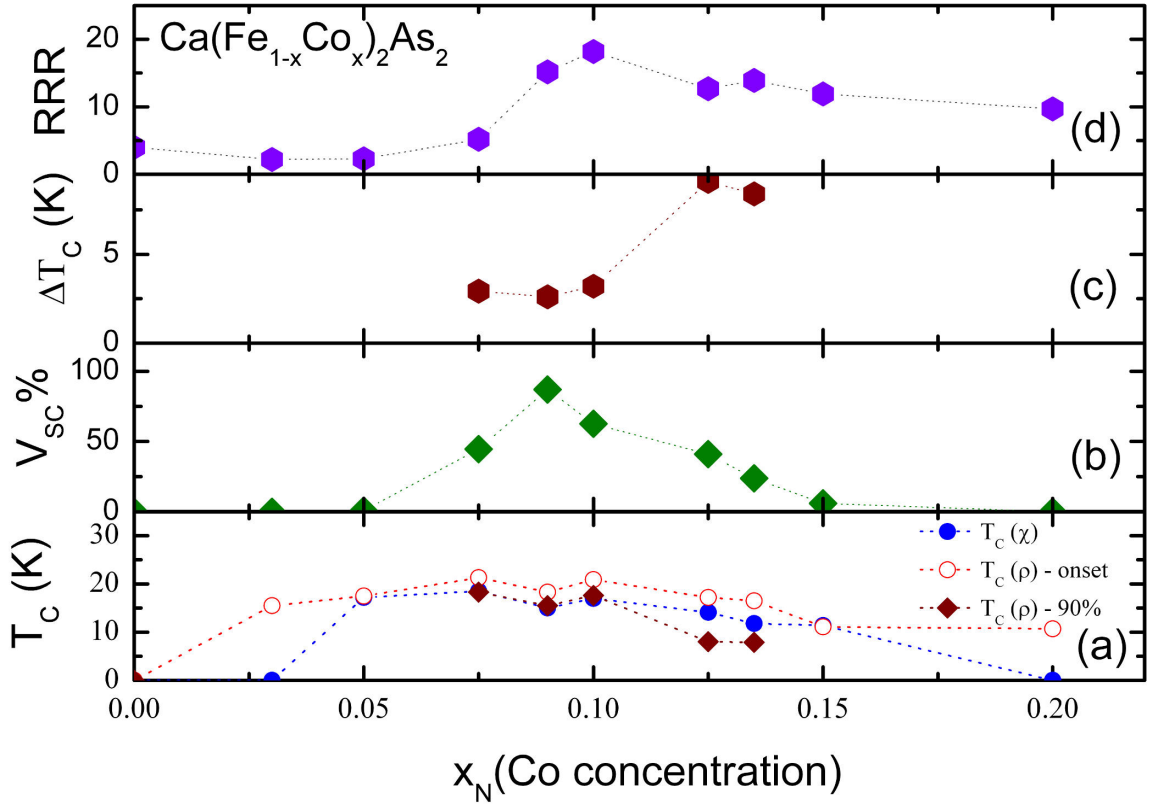

FIG. 13. (Color online) Evolution of superconducting state parameters in Ca(Fe$_{1-x}$Co$_x$)$_2$As$_2$ with Co concentration x. (a) Superconducting transition temperatures determined from the susceptibility and the transport data; (b) superconducting volume fraction determined from the magnetic susceptibility data; (c) variation of the width in temperature of the resistive superconducting transition T$_C$; (d) variation of the residual resistivity ratio with Co concentration. All lines are guides to the eye.